\documentclass[intlimits,twoside,a4paper]{article}

\usepackage{amsmath,amssymb}
\usepackage{graphicx}

\usepackage[T2A]{fontenc}
\usepackage[cp1251]{inputenc}

\usepackage{grffile}

\usepackage[eqsecnum]{cmpj2}

\issue{2013}{16}{3}{31704}
\doinumber{10.5488/CMP.16.31704}

\articletype{Proceedings Paper}

\title[Polarization switching in TGS and SBN]%
{Comparison of polarization switching in ferroelectric TGS and relaxor SBN crystals}

\author[K. Matyjasek, M. Or\l{}owski]{K. Matyjasek, M. Or\l{}owski}
\address{Institute of Physics, Faculty of Mechanical Engineering and Mechatronics, West Pomeranian University of Technology, Al. Piast\'{o}w 48, 70--310 Szczecin, Poland}

\date{Received February 10, 2013}
\authorcopyright{K. Matyjasek, M. Or\l{}owski, 2013}

\begin{document}

\maketitle

\begin{abstract}

The comparative experimental analysis of polarization reversal kinetics in conventional homogeneous triglycine sulfate ((NH$_{2}$CH$_{2}$COOH)$_{3}\cdot$H$_{2}$SO$_{4}$; TGS) and relaxor strontium barium niobate (Sr$_{0.61}$Ba$_{0.39}$Nb$_{2}$O$_{6}$; SBN) crystals have been performed  in a broad range of measurement conditions. The experimental data have been collected from microscopic observation of the domain structure, switching current and $D-E$ hysteresis loop registration. The hysteresis loop and dielectric spectra have a strong link to the configuration of ferroelectric microdomains. The domain structure dynamics was examined by  the nematic liquid crystal (NLC) method.
\keywords polarization switching, ferroelectric domain,  hysteresis loop, TGS, relaxor, SBN
\pacs 77.80.Dj, 77.80.Fm, 77.84.Dy
\end{abstract}

\section{Introduction}

Most of the ferroelectrics applications (including nonvolatile memories and a broad range of electronic, optical and acoustic devices) require control of the local polarization state. Thus, it is critical to understand the process of polarization switching \cite{a}. In this report, the effects of microstructure of domains (point defects) on the ferroelectric properties are established experimentally in rather different classes of ferroelectric materials, in well studied conventional ferroelectrics TGS and SBN relaxor crystals. Despite its complicated chemical and crystallographic form, TGS exhibit a wide range of features indispensable for understanding  very basic features of the mechanism of polarization switching \cite{b}. Compared to conventional TGS, a unique property of relaxor SBN is the appearance of a very broad and frequency-dependent dielectric anomaly near the ferroelectric phase transition \cite{c}. The dielectric properties of relaxors can be attributed to the development of quenched random fields associated with a compositional/structural disorder \cite{d}. Microscopic studies of the switching process in conjunction with electric measurements allowed us to establish a relationship between local properties of the domain dynamics and macroscopic responses such as polarization hysteresis loop , switching current and dielectric permittivity measurements. The microscopic features of 180 degree domain wall dynamics, in low electric fields, were investigated by NLC  method. The switching current transients have been analyzed by Kolmogorov-Avrami-Ishibashi (KAI) model based on the classical theory of nucleation and  domain growth \cite{e}.

\section{Experimental methodology}

To observe the optically indistinguishable 180$^{\circ}$ domain walls in the both crystals,  the NLC mixture of p-methoxybenzylidene-p-n-butylaniline (MBBA)  and pethoxybenzylidene-p-n-butylaniline  (EBBA)  was used.  Plateled-shaped samples of TGS and SBN  were cut perpendicularly to the  polar axis. A cover glass coated with a conducting layer of SnO$_{2}$ was used to observe the domain pattern evolution during polarization reversal in an electric field.  The regions, where the domain reorientation still occurs, look darker  because in these regions an electrohydrodynamic instability, particularly dynamic scattering, takes place \cite{f}. Hysteresis loops ($D-E$ dependence) were recorded with a modified Sawyer-Tower circuit by applying an ac-field of 50~Hz. The switching currents were measured by  applying square-wave electric pulses (two positive pulses followed by two negative ones) amplified with a Kepco bipolar amplifier. Dielectric permittivity dependence on temperature and frequency was measured by HP 4284A LCR meter. The electrical measurements were carried out with air-drying silver paste as electrodes.

\section{Results and discussion}

\begin{figure}[!b]
\centerline{
\includegraphics[width=0.65\textwidth]{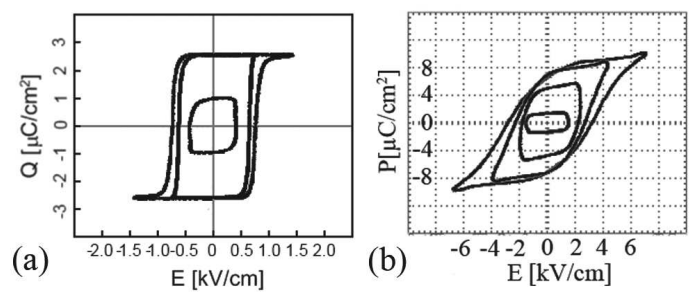}
}
\caption{The family of hysteresis loop obtained by applying ac-electric field of 50~Hz at room temperature for (a) TGS and (b) SBN crystal sample.} \label{Fig. 1}
\end{figure}

The ability to reverse their polarization state in ferroelectric materials under the application of an electric field determines their characteristic electric displacement~-- electric field ($D-E$) hysteresis loop. Figure~\ref{Fig. 1} shows the shape of hysteresis loop (H-L) for TGS [figure~\ref{Fig. 1}~(a)] and SBN [figure~\ref{Fig. 1}~(b)].

The measurement of H-L in a broad range of ac-electric field shows that the relaxor SBN does not have a definite coercive field. At a low electric field,  H-L has a nearly square shape relative to $D-E$ axis, similar to the one observed for TGS crystal, indicating an abrupt change of the polarization orientation. At higher $E$, the H-L for SBN becomes more slanted. It means that the slowly switching regions being initially frozen, become activated and participate in the polarization reversal  process.

\begin{figure}[!t]
\centerline{
\includegraphics[width=0.9\textwidth]{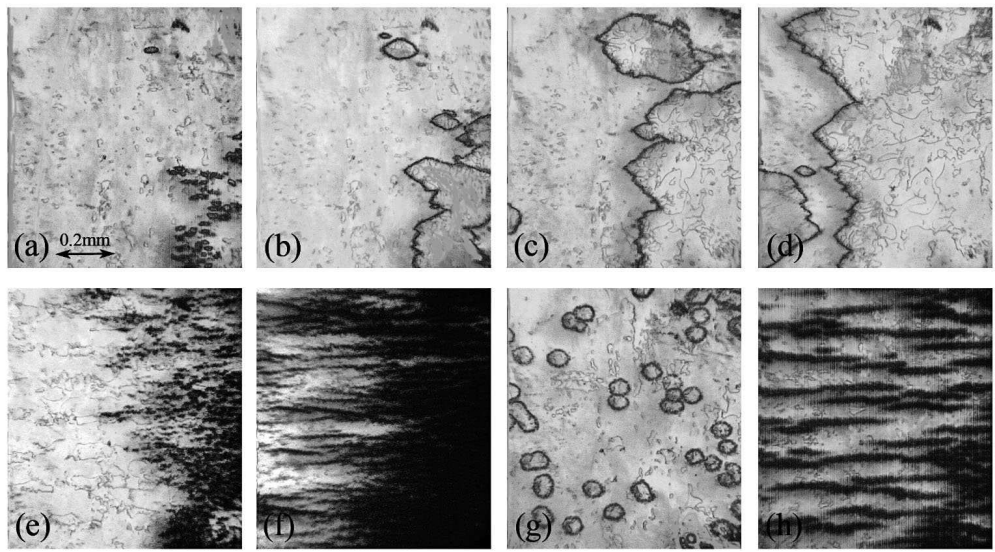}
}
\caption{\looseness=-1 The domain pattern evolution observed  in TGS crystal sample during switching process in  negative electric field of 0.4~kVcm$^{-1}$ (applied at $t=0$). Time from  the moment of applying $E$ in [s] (a) 0.2;  (b) 0.4;  (c) 0.8;  (d) 1.8; Distribution of domains in: negative $E$ in [kVcm$^{-1}$]  (e) 0.5; (f) 0.7; positive $E$  (g) 0.4; (h)~0.7.} \label{Fig. 2}
\end{figure}

The presence and the change of configuration of the domains determines the polarization hysteresis. An essential difference between the domain dynamics is observed in normal homogeneous ferroelectric  TGS and relaxor SBN crystals. In both type of crystals, the nucleation of ferroelectric domains takes place when the applied field exceeds a critical nucleation threshold (coercive field), which is subject to regional variation for defects. Figure~\ref{Fig. 2}~(a)--(d) shows a series of video frames illustrating the domain pattern evolution in TGS crystal starting with a single domain state in the negative  electric field of  0.4~kVcm$^{-1}$.

This field is high enough to complete the domain switching in the entire volume of the crystal sample. The polarization switching takes place through  inhomogeneous nucleation process and anisotropic growth of  the domains. The growth of the existing domains is more favorable than the creation of new ones. Thus, the nucleated domains expand with little or no resistance under an electric field and start to coalesce into large ones accompanied by a decrease of the domain density. The region where intensive nucleation takes place becomes larger on increasing the electric field [figure~\ref{Fig. 2}~(e)]. In a high electric field, the density of the domain nuclei arranged in rows and the number of rows per unit length increase, so that optically they become indistinguishable [see figure~\ref{Fig. 2}~(f)]. The inhomogeneous distribution of the domain nuclei during switching is a result of non-uniform internal field distribution, which has been attributed to the defects in the bulk of the crystal. A more homogeneous distribution of the domains during switching has been observed in positive electric fields , as shown in figures~\ref{Fig. 2}~(g) and~\ref{Fig. 2}~(h).

\begin{figure}[!b]
\centerline{
\includegraphics[width=0.9\textwidth]{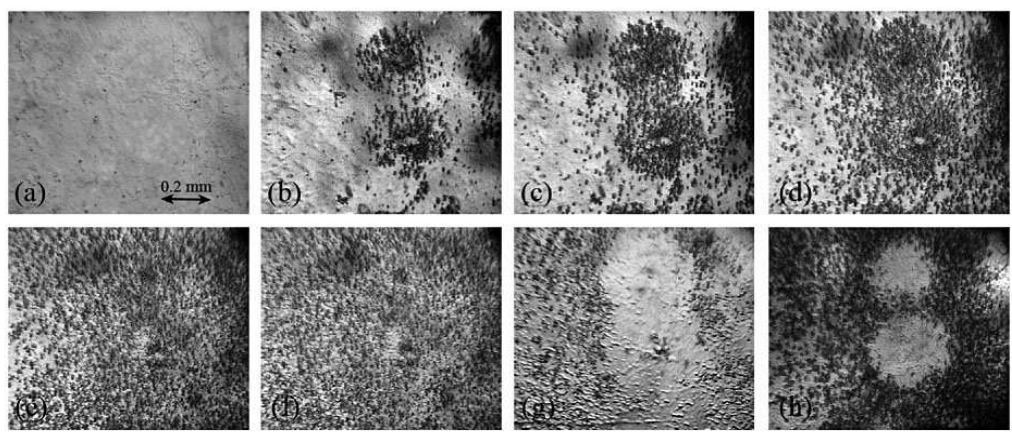}
}
\caption{Domain pattern evolution observed in SBN crystal sample during switching in the positive electric field of 2.8~kVcm$^{-1}$. (a)~--- initial single domain state. Time from the moment of applying $E$ in [s]; (b)~--- 0.2; (c)~--- 0.4; (d)~---  0.6; (e)~--- 0.8; (f)~--- 1; (g)~--- 3.8; (h)~--- shows the domain pattern obtained after applying the negative electric field  of 2.8~kVcm$^{-1}$.}
\label{Fig. 3}
\end{figure}

Evolution of the domain structure is more complex in SBN relaxor crystal.  Figure~\ref{Fig. 3} illustrates the domain pattern evolution in SBN, starting with a single  domain state, in the electric field of 2.8~kVcm$^{-1}$.

\looseness=-1This  field is high enough to complete the micro-scale domain switching in the entire volume of the sample, because a further increase of the electric field does not induce any domain switching. Unlike the TGS crystals, in relaxor SBN  the creation of new domains is more favorable than their growth.  The nucleation process continues to take place during almost the whole polarization reversal process, at a constant electric field.  The specific mechanism of nucleation in SBN could be interpreted in terms of a wide distribution of activation energies for nucleation resulting from the local structural irregularities~\cite{g, h}.

\begin{figure}[!t]
\centerline{
\includegraphics[width=0.9\textwidth]{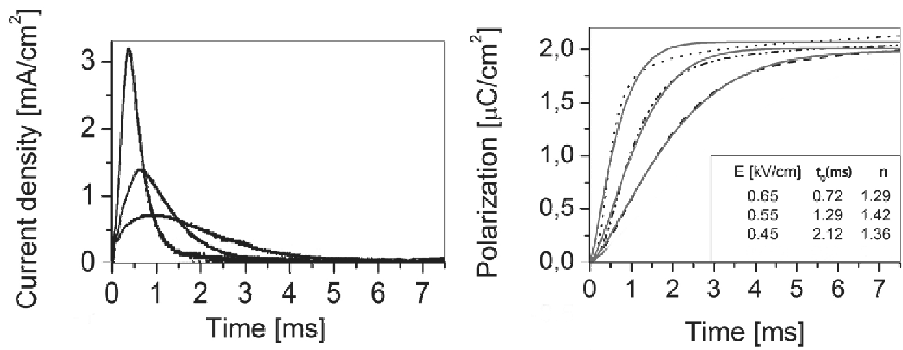}
}
\caption{Switching currents (a) and switching polarization (b) versus time for TGS crystal sample. The polarization data were fitted with Avrami function (solid curves).}
\label{Fig. 4}
\end{figure}

\begin{figure}[!b]
\centerline{
\includegraphics[width=0.65\textwidth]{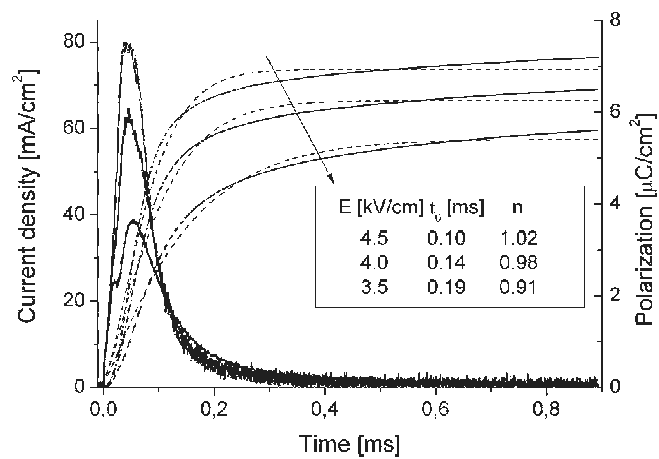}
}
\caption{Switching currents and switched polarization versus time for SBN crystal sample. The polarization data were fitted with Avrami function (dashed curves).}
\label{Fig. 5}
\end{figure}

The fast polarization switching process was investigated by measuring the switching currents in response to the square wave electric pulses. The switching current $i(t)$ was obtained by subtracting  the non-switching current from a full current. The rate of polarization switching at constant $E$ can be formed by integration of $i(t)$ from $t=0$ to the instant $t$. The results for TGS  are presented in figure~\ref{Fig. 4} and for SBN  in figure~\ref{Fig. 5}.

For TGS crystal, the saturated polarization is independent of the applied field. In relaxor SBN crystal, the switched polarization increases with the electric field strength, but does not saturate. It means that there are slowly switching regions that do not contribute to the switching current signals even in $E > E_\mathrm{c}$  (coercive field).  This should be related to the pinning effect of the domain walls resulting from a disorder structure of SBN crystal \cite{i}.

The traditional model used in describing the switching kinetics of ferroelectrics, called the Kolmo\-gorov-Avrami-Ishibashi (KAI) model, is based on the classical statistical theory of nucleation and unrestricted domain growth \cite{e}. Microscopic observation has revealed that, in the  high electric field range in which electrical measurements were carried out, the density of the domain nuclei is so high that the linear dimension of the sample is much larger than the distance between the nuclei, as it was assumed in the  idealized KAI model. The KAI theory gives the polarization change $P(t)$ (called Avrami function)  as $P(t)=P_{0}\left\{1-\exp\left[-\left(t/t_{0}\right)^{n}\right]\right\}$,  where $n$ and $t_{0}$ are the effective dimensionality and characteristic time, respectively, and $P_{0}$ is the switchable polarization. The effective dimensionality $n$ is related to the actual growth dimension $d$ of the domain walls and the mechanisms for nucleation. One-dimensional growth  ($d = 1$) implies plate-like domains with the walls moving in one direction perpendicular to the ferroelectric axis. Two-dimensional growth ($d = 2$) occurs when the nuclei are considered to be cylinders [as shown in figure~\ref{Fig. 2}~(g)]. The exponent $n$ depends on the assumed nucleation scenario. The constant nucleation rate corresponds to the exponent $n=d +1$, whereas the case of one-step nucleation leads to $n=d$. In figures~\ref{Fig. 4} and \ref{Fig. 5}, the experimental data were fitted by  the Avrami function. The results for TGS crystal show that the effective dimension $n$ is lower than 2 and the characteristic time $t_{0}$ decreases with an increasing field. The switching kinetics is in reasonable agreement with that predicted theoretically for a continuous nucleation case, indicating a one-dimensional growth of domains. Microscopic observations [see figure~\ref{Fig. 2}~(h)] revealed that a one-dimensional growth in the high electric field range is a result of a particular distribution of the domain nuclei arranged in rows.  From the image analysis of the domain growth pattern in TGS, it has been found that the nucleation rates (especially in the high field range), as well as the velocities of domain walls do not stay constant throughout the switching process. In such a case, the non-integer value of dimensionality $n$ is typically obtained  \cite{j}. The retardation behavior in such real physical conditions have been explained through polarization process with a broad distribution (Gaussian distribution) of characteristic domain growth times \cite{k}, or the Lorentzian distribution of logarithmic switching times \cite{l}.

Although fitting curves (Avrami function) have given a good fitting quality for SBN crystal (figure~\ref{Fig. 5}),  the KAI model is not applicable to the polarization reversal of SBN crystal. Note that the  $n$ value less than 1 is not physically reasonable according to the KAI model, since the growth dimensionality could never be less than 1. A fundamental difference in the switching process is revealed in SBN, and may be accounted for by  the slow inhomogeneous domain growth in the presence of random pinning fields characteristic of relaxor ferroelectrics \cite{d}.

\begin{figure}[!h]
\centerline{
\includegraphics[width=0.9\textwidth]{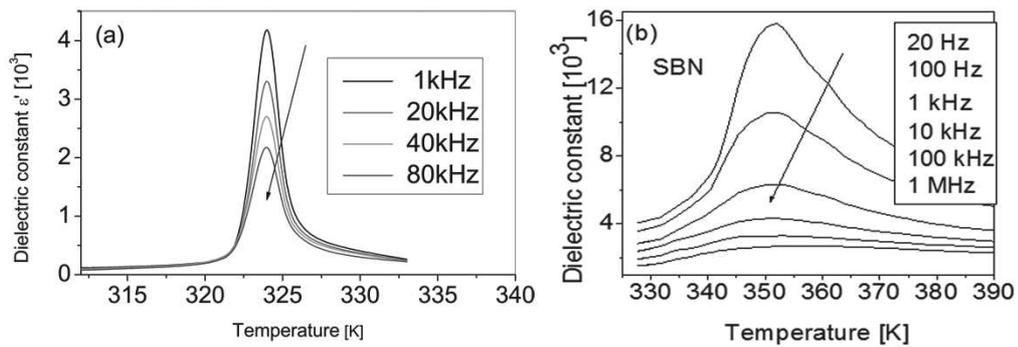}
}
\caption{Temperature dependences of dielectric constants measured at different frequencies for: (a)~--- TGS; (b)~--- SBN crystal sample.}
\label{Fig. 6}
\end{figure}

The domain walls have a considerable effect on the total dielectric response of  ferroelectric materials \cite{m}. Dielectric constants for TGS and SBN samples were measured versus temperature and the essential results are presented in figure~\ref{Fig. 6}.

The dielectric constant of the TGS exhibits a sharp, narrow peak at the phase transition temperature $T_\mathrm{c}$, and the width at half maximum is $\sim2 $~K.   By contrast, the relaxor SBN crystal exhibits very broad and frequency-dependent dielectric anomaly, and the width at half maximum is $\sim 20\div40$~K. The compositional disorder in SBN could play an important role on the phase transition broadening assuming the formation of the  polar regions with locally different Curie temperature $T_\mathrm{C}$ \cite{n}. Note that a nanopolar structure and local ferroelectricity have been revealed by piezoresponse force microcopy technique well beyond the phase transition temperature in SBN crystal \cite{o}.

\section{Conclusions}

The results demonstrate important differences in the polarization switching mechanism in conventional ferroelectric TGS  and relaxor SBN crystals.  It was shown that the domain growth is significant for TGS. In relaxor SBN crystal, the polarization reversal takes place mostly by nucleation of the domains. Thus, the KAI model of switching , based on statistics of domain coalescence,  can be applied only to  TGS crystals. The observed complexity of the domain structure  in relaxor SBN crystal could be understood in terms of a slow inhomogeneous domain growth in the presence of random pinning fields, which give rise to decelerated dynamics of domain walls.

\section*{Acknowledgements}

The authors would like to thank Dr. L. Ivleva from Russian Academy of Sciences in Moscow for providing SBN samples.



\begin{thebibliography}{99}

\bibitem{a}Soergel E.,  Appl. Phys. B, 2005, \textbf{81}, 729; \doi{10.1007/s00340-005-1989-9}.

\bibitem{b}Nakatani N., Ferroelectrics, 2011, \textbf{413}, 238; \doi{10.1080/00150193.2011.554269}.

\bibitem{c}Dec J., Kleemann W., Woike Th., Pankrath R., Eur. Phys. J. B, 2000, \textbf{14}, 627; \doi{10.1007/s100510051071}.

\bibitem{d}Kleemann W., J. Mater. Sci., 2006, \textbf{41}, 129; \doi{10.1007/s10853-005-5954-0}.

\bibitem{e}Ishibashi Y., Takagi Y., J. Phys. Soc. Jpn., 1971, \textbf{31}, 506; \doi{10.1143/JPSJ.31.506}.

\bibitem{f}Tikhomirova N.A.,  Dontsova L.J., Pikin S.A., Shuvalov L.A., JETP Lett., 1979, \textbf{29}, 34.

\bibitem{g}Gladkii V.V., Kirikov V.A., Volk T.R., Ivleva L.I., Ferroelectrics, 2003, \textbf{285}, 275; \doi{10.1080/00150190390206112}.

\bibitem{h}Matyjasek K., Wolska K., Rogowski R. Z, Kaczmarek S. M., Ivleva L. I., Ferroelectrics, 2011, \textbf{413}, 311;  \\ \doi{10.1080/00150193.2011.531212}.

\bibitem{i}Glass A.M., J. Appl. Phys., 1969 , \textbf{40}, 4699; \doi{10.1063/1.1657277}.

\bibitem{j}Matyjasek K., J. Phys. D: Appl. Phys., 2001, \textbf{34}, 2211; \doi{10.1088/0022-3727/34/14/317}.

\bibitem{k}Rogowski R. Z., Matyjasek K., Jakubas R., J. Phys. D: Appl. Phys., 2005, \textbf{38}, 4145; \doi{10.1088/0022-3727/38/23/001}.

\bibitem{l}Jo J.Y., Han H.S.,  Yoon J.G.,  Song T.K., Kim S.H., Noh T.W.,  Phys. Rev. Lett., 2007, \textbf{99}, 267602; \\ \doi{10.1103/PhysRevLett.99.267602}.

\bibitem{m}Lines M.E., Glass A.M., Principles and Applications of Ferroelectrics and Related Materials, Oxford University Press, Oxford, 2001.

\bibitem{n} Bokov A.A. , Ye, Z.-G., J. Mater. Sci., 2006, \textbf{41}, 31; \doi{10.1007/s10853-005-5915-7}.

\bibitem{o} Liu X.Y.,  Liu Y.M., Takekawa S.,  Kitamura K., Ohuchi F.S., Li J.Y., J. Appl. Phys., 2009,  \textbf{106}, 124106; \\ \doi{10.1063/1.3273481}.



\end{thebibliography}

\ukrainianpart

\title{Порівняння перемикання поляризації в сегнетоелектричному TGS і
релаксорному SBN кристалах}

\author[К. Матиясек, М. Орловскі]{К. Матиясек, М. Орловскі}
\address{Інститут фізики, факультет механічної інженерії і
мехатроніки, \\
Західно-померанський технологічний університет, Щецін, Польща}

 \makeukrtitle
\begin{abstract}

Здійснено порівняльний експериментальний аналіз  оборотної кінетики
поляризації в стандартному  однорідному кристалі тригліцин сульфату
((NH$_{2}$CH$_{2}$COOH)$_{3}\cdot$H$_{2}$SO$_{4}$; TGS) і
релаксорному кристалі стронцій барієвого ніобату
(Sr$_{0.61}$Ba$_{0.39}$Nb$_{2}$O$_{6}$; SBN) в широкій області
вимірювальних умов. Експериментальні дані були зібрані з
мікроскопічного спостереження доменної структури, струмовового
перемикання і  реєстрації петлі гістерезису $D-E$. Петля
гістерезису і діелектричні спектри мають сильний зв'язок з
конфігурацією сегнетоелектричних мікродоменів. Динаміка доменної
структури вивчалася методом нематичного рідкого кристалу (NLC).

\keywords перемикання поляризації, сегнетоелектричний домен, петля
гістерезису, TGS, релаксор, SBN
\end{abstract}\end{document}